\documentstyle[preprint,tighten,aps]{revtex}

\begin{document}

\vspace*{2.5cm}
{\Large\bf Relational evolution of the degrees of freedom of generally 
covariant quantum theories}

{\bf Merced Montesinos\footnote{ 
        Departamento de F\'{\i}sica, Centro de Investigaci\'on y de
        Estudios Avanzados del I.P.N.,\\
        Av. I.P.N. No. 2508, 07000 Ciudad de M\'exico, M\'exico.
        e-mail: merced@fis.cinvestav.mx}}

\begin{center}
\begin{minipage}[t]{9.5cm}\footnotesize
\hrulefill

We study the classical and quantum dynamics of generally covariant 
theories with vanishing Hamiltonian and with a finite number of degrees of 
freedom. In particular, the geometric meaning of the full solution of the 
relational evolution of the degrees of freedom is displayed, which means the 
determination of the total number of evolving constants of motion 
required. Also a method to find evolving constants is proposed. The 
generalized Heinsenberg picture needs $M$ time variables, as opposed to the 
Heisenberg picture of standard quantum mechanics where one time variable 
$t$ is enough. As an application, we study the parameterized harmonic 
oscillator and the $SL(2,R)$ model with one physical degree of freedom that 
mimics the constraint structure of general relativity where a Schr\"odinger 
equation emerges in its quantum dynamics. 

\hrulefill
\end{minipage}
\end{center}


\section*{Introduction}
Describing {\it evolution} of the degrees of freedom of generally covariant 
theories is an unsolved puzzle, and constitutes one of the challenges 
of the human thinking of our time. The study of generally covariant theories 
has been motivated by general relativity, which has this peculiar 
property (see for instance \cite{Stachel1989}). In gravity, general 
covariance means the theory is diffeomorphism invariant, and this 
symmetry of gravity implies at the Hamiltonian level that the theory has 
not a genuine Hamiltonian for describing the evolution of the degrees of 
freedom of the gravitational field, rather, dynamics is gauge; generated 
by the first class constraints of the theory. This is the so-called problem 
of time in classical general relativity \cite{Rovelli1999}.

On the other hand, if the classical regime of general relativity is only a 
limit, which emerges in a suitable way from its fundamental quantum 
behavior , then the theory is in trouble. Our standard methods of 
quantization crash and do not apply to the particular physical situation 
raised by general relativity. Standard quantization methods in field theory 
are background-dependent, quantum gravity needs a background independent 
procedure in its quantization. So, how to make compatible the symmetry 
of general relativity with the quantum theory? what is the meaning of 
evolution in quantum gravity? Loop quantum gravity answers the first 
question, because the quantization of the gravitational field is carried 
out in a background independent way \cite{Rovelli:1998}. With respect to 
the second, it remains as an open question. Among the several proposals 
for describing the evolution of the degrees of freedom of generally 
covariant theories, I find Rovelli's proposal as one of the most creative 
ones \cite{Rovelli1990,Rovelli1991a,Rovelli1991}. 

Here, following the spirit of relationism, which is the heart 
in Rovelli's point of view, we analyse the `problem of time' in generally 
covariant theories with vanishing Hamiltonian and with a finite number 
$D$ of degrees of freedom. To obtain the relations involving the 
coordinates and momenta of the unreduced phase space $\gamma_{ex}$ with the 
physical states that label the points of the physical phase 
space $\gamma_{ph}$, we need to start from the embedding equations which 
give the dependence of the coordinates and momenta with respect to the $M$ 
time variables $t^m$ as well as the $2D $ physical states 
$({\tilde q}^a , {\tilde p}_a)$. These equations constitute the classical 
version of the generalized Heisenberg picture, which arises when these 
equations are promoted to quantum operators in the reduced Hilbert space 
of the theory. By plugging the expressions of the time variables $t^m$ in 
terms the original canonical variables into the expressions of coordinates 
and momenta, we get the full relational evolution of the phase space degrees 
of freedom for any physical state of ${\gamma_{ph}}$. This way of expressing 
the full solution of the dynamics of generally covariant theories 
constitutes the full set of evolving constants of motion required in their 
dynamics, and is displayed in Sect. I. In addition, an alternative 
mechanism which generates also the evolving constants of motion is 
proposed. Of course, we study also the quantum version of the evolving 
constants of motion. In sect. II, we analyze the parameterized harmonic 
oscillator (as an example of parameterized systems). In Sect. III, we 
continue the study of the $SL(2,R)$ model which constraint algebra mimics 
the algebra structure of general relativity. Due to the fact a Schr\"odinger 
equation emerges in its quantum dynamics, we compare the generalized 
Heisenberg picture (related with the evolving constants of motion) which 
needs $M$ time variables with the Schr\"odinger picture which singles out 
one time variable only. We also especulate on the classical limit generally 
covariant theories and its possible relation with the full set of evolving 
constants of motion. Our conclusions are summarized in Sect. IV.

\section{Relational classical and quantum dynamics}
{\it Classical dynamics}. The classical dynamics of a constrained theory 
with a finite number of degrees of freedom characterized by first class 
constraints is as follows \cite{Dirac}. The theory is obtained from 
the Hamiltonian form of the action
\begin{eqnarray}
S [ q^i , p_i , \lambda^m ]& = & \int_{\tau_1}^{\tau_2} d \tau \,\, 
\left \{ \frac{dq^i}{d\tau}p_i - \lambda^m C_m (q^i,p_i) \right \} \, ,
\end{eqnarray} 
which is invariant under arbitrary reparameterizations of the parameter 
$\tau$. Hence, $\tau$ is a non physical coordinate time. The unreduced phase 
space $\gamma_{ex}$ is coordinated by the canonical pairs 
$(q^i , p_i)$ ;$ i=1,2,...,N$. The canonical 2-form on $\gamma_{ex}$ is 
$\omega_{ex} =dp_i \wedge d q^i$. Thus, $(\gamma , \omega)$ is a symplectic 
space. The variation of the action $S [ q^i , p_i , \lambda^m ]$ with 
respect to the canonical coordinates $q^i , p_i$ gives the equations of 
motion
\begin{eqnarray}
\frac{dq^i}{d\tau} & = &  \lambda^m 
\frac{\partial C_m (q^i ,p_i)}{\partial p_i} \, , \nonumber\\
\frac{dp_i}{d\tau} & = & - \lambda^m 
\frac{\partial C_m (q^i ,p_i)}{\partial q^i} \, ,
\end{eqnarray}
while the variation of the action with respect to the Lagrange multipliers 
$\lambda^m$ gives the constraint equations
\begin{eqnarray}
C_m & = & C_m (q^i ,p_i) = 0\, , \quad m=1,2...,M  \, .
\label{firstclass}
\end{eqnarray}
The variation of the action has been done under the standard boundary 
conditions $q^i (\tau_s) = q^i_s$; $s=1,2$, namely, the allowed paths are 
those with fixed values for the configuration variables at the boundary 
points $\tau_s$. The boundary conditions can be changed and thus to modify 
the action by suitable boundary terms to allow the gauge symmetry generated 
by the constraints \cite{Henneaux1992}. The constraints generate Hamiltonian 
vector fields $X_{dC_m}$, which are tangent vectors to the constraint 
surface, given by
\begin{eqnarray}
X_{dC_m} & = & -  \frac{\partial C_m}{\partial p_i} 
\frac{\partial}{\partial q^i} + 
\frac{\partial C_m}{\partial q^i} 
\frac{\partial}{\partial p_i} \, . \label{HVF}
\end{eqnarray}
More important, the integral curves of these Hamiltonian vectors fields 
constitute the gauge submanifold or the orbits of the constraint
surface, and the dynamics of the system with respect to $\tau$ is the 
unfolding of this gauge symmetry, i.e., dynamics is gauge.

The first class constraints satisfy, in general, a non Lie algebra
\begin{eqnarray}
\{ C_m , C_n \} & = & C_{mn}\,^l (q^i , p_i) C_l \, ,
\end{eqnarray}
and the number of independent physical degrees of freedom of the theory 
is $D=N-M$. The constraint surface defined by the constraint equations 
(\ref{firstclass}) is a $(2D+M)$-dimensional manifold. The constraint 
surface can be parameterized by the set of independent coordinates 
$({\tilde q}^a , {\tilde p}_a , t^m)$, where 
$({\tilde q}^a , {\tilde p}_a), a=1,2,...,D$ are (local) canonical 
variables which coordinatize the open sets of the physical phase 
space $\gamma_{ph}$ of the theory, and $t^m , m=1,2,...,M$ 
coordinatize the orbits, i.e., the gauge submanifold 
of the constraint surface generated by the first class constraints. Notice 
that the canonical coordinates $q_0\,^{\alpha}$ , and $p_{0\alpha}$ 
{\it are} the physical observables of the system. Of course, they satisfy 
$\{ {\tilde q}^a, {\tilde p}_b \}=\delta^a_b$ on the physical 
phase space, and the symplectic form on $\gamma_{ph}$ in these coordinates 
is $\omega_{ph}= d{\tilde p}_a \wedge d {\tilde q}^a $. Therefore, the 
general solution of the dynamics of the constrained theory can be expressed as
\begin{eqnarray}
q^i & = & q^i (t^m; {\tilde q}^a, {\tilde p}_a ) \, , \label{ECI}\\
p_i & = & p_i (t^m; {\tilde q}^a , {\tilde p}_a ) \, \label{ECII}.
\end{eqnarray}
It is important to emphasize that that such dependence is local. For 
instance, in the case in which the physical phase space $\gamma_{ph}$ 
is compact, a finite set of physical observables 
$( {\tilde q}^a , {\tilde p}_a )$ is needed to coordinate the open sets of 
$\gamma_{ph}$ due to its compactness. So, the constraint surface looks like 
a `fibre bundle' $P(\gamma_{ph},\mbox{Orbits})$, the constraint surface 
being the total space $P$, the physical phase space $\gamma_{ph}$ being 
the base space, and the orbits being the fibers of the bundle. In the 
generic case, $P(\gamma_{ph},\mbox{Orbits})$ is locally trivial. This 
means that non global gauge condition is allowed in general, and that local 
gauge conditions associated with each open set of the physical phase space 
can be specified only.

At the same time, the full solution of the theory implies to give the 
dependence of the physical observables ${\tilde q}^a$, and ${\tilde p}_a$ of 
the system in terms of the coordinates of the unreduced phase space
\begin{eqnarray}
{\tilde q}^a & = & {\tilde q}^a (q^i, p_i) \, \label{OI},\\
{\tilde p}_a & = & {\tilde p}_a (q^i, p_i) \, \label{OII},
\end{eqnarray}
as well as the orbit coordinates $t^m$
\begin{eqnarray}
t^m & = & t^m (q^i ,p_i)\, . \label{GI}
\end{eqnarray}
What these equations tell us is that one single internal time variable is 
not enough to describe the evolution of the system, rather, $M$ internal 
time variables are needed. In the way the full solution has been expressed 
in (\ref{ECI}), and (\ref{ECII}), these $M$ time variables are $t^m$, 
$m=1,2,...,M$. Notice also that these $M$ time variables are {\it internal} 
clocks, given by (\ref{GI}). One of the properties of these internal clocks 
$t^m$ is that they do {\it not} run taking increasing values of $t^m$ when 
time goes on. In fact, they can run `forward' and `backward' depending on the
values of the coordinates and momenta the system is reaching through 
Eq. (\ref{GI}). Other property is that these clocks can run with different 
`speeds' for the same reason. So, the meaning of time that arise in 
generally covariant theories is completely different with respect to the 
monotonous function we are familiarized with.  In \cite{Goldberg91} 
was showed that Eqs. (\ref{ECI}) and (\ref{ECII}) can be obtained by a 
combination of a canonical transformation plus Hamilton-Jacobi 
techniques. That approach implies the modification of the original set of 
first class constraints. This can be done, but it is not 
necessary in principle. Moreover, the full solution requires (\ref{GI}) 
(missing in Ref. \cite{Goldberg91}) as we have seen, and more important, it 
is the combination of Eqs. (\ref{ECI}), and (\ref{ECII}) with Eq. (\ref{GI}) 
which leads to the relational description of the dynamics of the system, as 
we will see it later on.

It is worth to mention the relationship between the time variables 
$t^m$ and the full gauge transformation generated by the first class 
constraints $C_m$. Assuming that the full gauge transformation of the 
original canonical variables is given by
\begin{eqnarray}
{q'}^i & = & {q'}^i (q^i , p_i , \alpha^m (\tau)) \, , \nonumber\\
{p'}_i & = & {p'}_i (q^i , p_i , \alpha^m (\tau)) \, , \label{Timegauge}
\end{eqnarray}
with $\alpha^m$ the $m$ gauge parameters involved in the gauge 
transformation. Then, by plugging (\ref{ECI}) and (\ref{ECII}) into the 
right hand side of (\ref{Timegauge}), we get
\begin{eqnarray}
{q'}^i & = & {q}^i ({t'}^m ; {\tilde q}^a , {\tilde p}_a ) \, ,\\
{p'}_i & = & {p}_i ({t'}^m ; {\tilde q}^a , {\tilde p}_a ) \, , 
\end{eqnarray} 
where the functional dependence in the right-hand side of the above 
equations is exactly the same as that given by (\ref{ECI})-(\ref{ECII}) but 
with 
\begin{eqnarray}
{t'}^m & = & {t'}^m (t^m , \alpha^m )\, ,
\end{eqnarray}
which relates the time variables $t^m$ after and before of any finite gauge 
transformation of the canonical variables (\ref{Timegauge}).

The map 
$\phi : P(\gamma_{ph},\mbox{Orbits}) \rightarrow \gamma_{ex}, 
\phi ({\tilde q}^a , {\tilde p}_a , t^m) \mapsto 
(q^i (t^m; {\tilde q}^a , {\tilde p}_a ), 
p_i (t^m; {\tilde q}^a , {\tilde p}_a ))$ allows us to define on the 
constraint surface 
$P(\gamma_{ph},\mbox{Orbits})$ the pull back 
$\Omega = \phi^{\ast} \omega_{ex}$ of the canonical form 
$\omega_{ex} =dp_i \wedge dq^i$ on $\gamma_{ex}$, which is 
degenerate. Thus, the geometry of constrained systems involves 
three spaces: the unconstrained phase space $(\gamma_{ex}, \omega_{ex})$, 
the constraint surface $(P(\gamma_{ph},\mbox{Orbits}), \Omega = \phi^{\ast} 
\omega_{ex})$, and the physical phase space 
$(\gamma_{ph}, \omega_{ph})$. The map that connects the constraint surface 
and the physical phase space is the projection 
$\pi: P(\gamma_{ph},\mbox{Orbits}) \rightarrow \gamma_{ph}$, 
$\pi ({\tilde q}^a , {\tilde p}_a , t^m) \mapsto ({\tilde q}^a , 
{\tilde p}_a )$. Due to the fact that the Hamiltonian vector fields 
(\ref{HVF}) are tangent vectors to the orbits, they can be expressed in 
terms of the local coordinates ${\tilde q}^a , {\tilde p}_a , t^m$ of 
the constraint surface as
\begin{eqnarray}
X_{dC_m} & = & \{ C_m , t^n \}({\tilde q}^a , {\tilde p}_a , t^m) 
\frac{\partial}{\partial t^n} \, .
\end{eqnarray}
The observables (\ref{OI}), (\ref{OII}), and the orbits (\ref{GI}) also 
generate Hamiltonian vectors fields, their restriction on the constraint 
surface are
\begin{eqnarray}
X_{d {\tilde q}^a } & = & \frac{\partial}{\partial {\tilde p}_a} + 
\{ {\tilde q}^a , t^n \} 
\frac{\partial}{\partial t^n} \, , \nonumber\\
X_{d {\tilde p}_a} & = & -\frac{\partial}{\partial {\tilde q}^a } + 
\{ {\tilde p}_a , t^n \} 
\frac{\partial}{\partial t^n} \, , \nonumber\\
X_{d t^m} & = & \{ t^m , {\tilde q}^a \}  
\frac{\partial}{\partial {\tilde q}^a } +
\{ t^m , {\tilde p}_a \}
\frac{\partial}{\partial {\tilde p}_a} + \{ t^m , t^n \}
\frac{\partial}{\partial t^n}\, ,
\end{eqnarray}
where it is understood that all the quantities are evaluated in the point 
$({\tilde q}^a , {\tilde p}_a , t^m)$. Thus, 
$\{ X_{d C_m} , X_{d {\tilde q}^a} , X_{ d {\tilde p}_a} \}$ is a basis, 
naturally adapted to the involved geometry, of the tangent space of the 
constraint surface. The vectors $X_{d C_m}$ play the role of vertical 
vectors because they have vanishing projection on the tangent space of 
$\gamma_{ph}$, $d\pi \left ( X_{d C_m}  \right) = 0 $.  
$X_{d {\tilde q}^a}$ , and $X_{ d {\tilde p}_a}$ are the horizontal lifts on 
the constraint surface of 
the coordinate basis on $\gamma_{ph}$, 
$d \pi \left ( X_{d {\tilde q}^a} \right ) = 
\frac{\partial}{\partial {\tilde p}_a}$ , 
$d \pi \left ( X_{d {\tilde p}_a} \right ) = 
- \frac{\partial}{\partial {\tilde q}^a}$. In summary, the solution of the 
dynamics of the constrained system means 
to specify Eqs. (\ref{ECI})-(\ref{GI}). This fact, rises a new problem: the 
problem of the meaning of {\it physical time} of generally covariant 
theories, i.e. the specification of an internal clock in the framework of the 
theory with respect to which to describe the evolution of the degrees of 
freedom of the theory in a {\it gauge invariant} way. Let us explain, the 
dynamics with respect to $\tau$ is given by
\begin{eqnarray}
q^i (\tau) & = & q^i (t^m (\tau); {\tilde q}^a, {\tilde p}_a ) \, , \\
p_i (\tau) & = & p_i (t^m (\tau); {\tilde q}^a , {\tilde p}_a ) \, ,
\end{eqnarray} 
for any physical state $({\tilde q}^a, {\tilde p}_a)$ of the system. So, 
this dynamics is non gauge-invariant, i.e., it depends on $\tau$. The 
question is, can we describe evolution of the system in a gauge invariant 
way? The answer is yes. At first sight, this sounds strange or impossible in 
a system with gauge freedom. To see how this can be achieved, let us plug the 
time variables (\ref{GI}) into the full solution (\ref{ECI}), and (\ref{ECII})
\begin{eqnarray}
q^i & = & q^i ( t^m (q^i , p_i ); {\tilde q}^a, {\tilde p}_a ) \, , 
\label{FSI} \\
p_i & = & p_i (t^m (q^i , p_i ); {\tilde q}^a, {\tilde p}_a ) \, .
\label{FSII}
\end{eqnarray} 
Last equations are very important, they relate the original phase space 
variables $q^i$, and $p_i$ with the physical states of the physical 
phase space $( {\tilde q}^a , {\tilde p}_a )$. These equations admite 
two, related, interpretations. First, they give the relational evolution of 
the coordinates  $q^i$ and the momenta $p_i$ for any fixed point 
$({\tilde q}^a, {\tilde p}_a )$ of the physical phase space, i.e., it is 
possible to choose M coordinates denoted by $q^m$ (or momenta $p_m$; or a 
combination of both) as `clocks' and describe the evolution of the 
remaining set of coordinates and momenta as functions of the $q^m$ for any 
physical state $({\tilde q}^a, {\tilde p}_a )$ of the system. Second, if we 
fix the values of this $M$ coordinates, say $q^m ={q}^{\ast m}$ then, the 
before mentioned expressions of coordinates and momenta give $M$-parameter 
families of physical observables defined on $\gamma_{ph}$, 
$q^{\ast m}$ being the parameters. Eqs. (\ref{FSI}), and (\ref{FSII}) 
are evolving constants of motion in the sense of Rovelli 
\cite{Rovelli1990,Rovelli1991a,Rovelli1991}. This concept captures the 
essence that the before mentioned observables (defined on $\gamma_{ph}$) 
describe the relational evolution of the coordinates $q^i$ and momenta 
$p_i$, and at the same time they are physical observables. 

Let us consider particular cases of (\ref{GI}), say
\begin{eqnarray}
t^m & = & q^m \, , \quad m =1,2,...,M \, ,
\end{eqnarray}
then (\ref{FSI}), and (\ref{FSII}) acquire the form
\begin{eqnarray}
q^m & = & q^m \, , \quad \quad \quad \quad\quad m=1,2,...,M \, ,\\
q^i & = & q^i (q^m ; {\tilde q}^a , {\tilde p}_a ) \, , \quad 
i= M+1, ... , N \, , \\
p_i & = & p_i (q^m ; {\tilde q}^a , {\tilde p}_a ) \, , \quad 
i= 1, ... , N \, .
\end{eqnarray}
Thus, the `clocks' are given by $q^m$ and last two pairs of equations are the 
evolving constants of motion involved. Other particular case is given by
\begin{eqnarray}
t^m & = & p_m \, , \quad m =1,2,...,M \, ,
\end{eqnarray}
and  (\ref{FSI}), and (\ref{FSII}) acquire the form
\begin{eqnarray}
q^i & = & q^i (p_m ; {\tilde q}^a , {\tilde p}_a )\, , 
\quad i=1,2,...,N\, , \\
p_m & = & p_m \, ,\quad \quad\quad\quad\quad m = 1,2,...,M \, , \\
p_i & = & p_i (p_m ; {\tilde q}^a , {\tilde p}_a ) \, , \quad 
i= M+1, ... , N \, .
\end{eqnarray}
In this case, the `clocks' are $p_m$ and last two pairs of equations are the 
evolving constants of motion required. 

As we have seen, the general relations involving the coordinates $q^i$ 
and momenta $p_i$ with the physical states $({\tilde q}^a , {\tilde p}_a)$ 
is given by ({\ref{FSI}), and (\ref{FSII}). The explicit 
form of (\ref{FSI}), and (\ref{FSII}) could be complicated for particular 
theories, but this fact would rise technical rather than conceptual 
difficulties (see \cite{Hajicek1991,Hartle1996} for the opposite 
viewpoint where the authors rise questions on interpretation, 
consistency, and the degree to which the resulting quantum theory emerging 
from the before classical dynamics coincide with, or generalizes, the usual 
non-relativistic theory). Thus, Eqs. (\ref{FSI}), and (\ref{FSII}) 
constitute the full set of evolving constants needed in the relational 
description of the dynamics of generally covariant theories with a finite 
number $D$ of degrees of freedom. The solution (\ref{FSI}), and (\ref{FSII}) 
sits in the spirit that in covariant theories there is non privileged 
observable with respect to which to describe evolution, and that only 
{\it relational evolution} makes sense. From this point of view, general 
covariance forces us to use relational evolution, namely, to describe 
the change of some variables of the system with respect to the others. This 
is the essence of relationism, which appears to be the natural language for 
describing the evolution of the degrees of freedom of generally covariant 
theories \cite{Rovelli1999,Rovelli1990,Rovelli1991a,Rovelli1991}.

In addition, in this paper, we propose an alternative mechanism to generate 
the evolving constants. This mechanism is essentially to compute the action 
of the Hamiltonian vector fields $X_{dC_m}$ on some evolving constant $E^1$
\begin{eqnarray}
X_{d C_m} (E) & =: & E^m \, .
\end{eqnarray}
The evolving function $E^1$ depends on the canonical coordinates of the 
unconstrained phase space $q^i$, and $p_i$ as well as on the canonical 
coordinates ${\tilde q}^a$, and ${\tilde p}_a$ of the physical phase 
space. Therefore, in the computation of the action of the Hamiltonian vector 
fields (\ref{HVF}) on the evolving function we can proceed in two 
ways. First, taking the observables ${\tilde q}^a$, and 
${\tilde p}_a$ constants in the dependence of the evolving function 
$E$. This can be done because ${\tilde q}^a$, and ${\tilde p}_a$ are 
constant along the orbits. b) Taking the explicit dependence of the physical 
observables in terms of the canonical variables of the unconstrained phase 
space given by (\ref{OI}), and (\ref{OII}). Of course, both approaches lead 
to the same results. The repeated application of the Hamiltonian vector 
fields on the new evolving constants $E^2$, $E^3$,.., gives another evolving 
constants, and so on until no new evolving constants are obtained, and the 
process ends. From the knowledge of the evolving constants and the 
expressions of the physical observables, the full solution of the dynamics 
of the system encoded in Eqs. (\ref{ECI})-(\ref{GI}) is obtained. 

{\it Quantum dynamics}. Let us begin with the quantum description of the 
system. We use the Dirac method. In the same way as in the classical 
dynamics we have three spaces $(\gamma_{ex}, \omega_{ex})$, 
$(P(\gamma_{ph}, \mbox{Orbits}) , \Omega)$, and 
$(\gamma_{ph}, \omega_{ph})$. In the quantum theory we have three Hilbert 
spaces; the unconstrained Hilbert space ${\cal H}$ or a suitable extension 
of it if the constraints have continuum spectrum, the physical Hilbert 
space ${\cal H}_{phys}$, and the reduced Hilbert space ${\cal H}_r$ obtained 
by projecting ${\cal H}_{phys}$. Suppose we have solved the quantum theory 
in a full way, i.e., we have the physical Hilbert space 
${\cal H}_{phys}$ of the theory. A general {\it physical } 
state $\mid \phi\rangle$ of the system is killed by all the constraints of 
the theory ${\widehat C}_m \mid \psi \rangle = \mid 0 \rangle$, and it is 
given by
\begin{eqnarray}
\mid \psi \rangle & = & \sum_{n_1, n_2,...,n_D} c_{n_1,n_2,...,n_D} 
\mid n_1, n_2,...,n_D\rangle \, . \label{abstract states}
\end{eqnarray}
in Dirac notation. Here, the physical states are labelled by the 
quantum numbers $n_a$, $a =  1,2,...,D$ which come from a 
complete set of commuting physical observables 
${\widehat O}_a$, $a =1,2,..., D$ of the system 
\begin{eqnarray}
{\widehat O}_a \mid n_1, n_2,...,n_D\rangle & = & 
O(n_a) \mid n_1, n_2,...,n_D \rangle \, .\nonumber
\end{eqnarray}
Of course, these quantum observables are combinations of the physical 
observables ${\widehat {\tilde q}}^a$, and 
${\widehat {\tilde p}}_a$. We have come to the heart of the problem, how to 
describe relational evolution in the quantum theory.

{\it Quantum evolving constants.} Let us see how the quantum version of the 
classical evolving constants looks. The idea is to search for a 
representation of the physical states (\ref{abstract states}) in the reduced 
Hilbert space associated with the physical phase space of the 
system. Explicitly
\begin{eqnarray}
\psi ({\tilde q}^a ) & = & \langle {\tilde q}^a \mid \psi \rangle = 
\sum_{n_1, n_2,...,n_D} c_{n_1,n_2,...,n_D} 
\langle {\tilde q}^a \mid n_1, n_2,...,n_D \rangle \, .
\end{eqnarray}
The inner product in the Hilbert space
\begin{eqnarray}
\langle \psi \mid \phi  \rangle &= & 
\int d\mu( {\tilde q}^a ) \,\, \psi^{\ast} ({\tilde q}^a ) 
\phi ({\tilde q}^a ) \, , \label{inner product}
\end{eqnarray}
can be determined with the condition that the operators 
${\widehat {\tilde q}}^a$, and ${\widehat {\tilde p}}_a$ be hermitian 
operators and with the implementation of the action of the 
operators ${\widehat {\tilde q}}^a$, ${\widehat {\tilde p}}_a$ on this 
Hilbert space. Notice also that is {\it always} possible to build 
creation and annihilation operators 
${\widehat a}_a = 
{\widehat {\tilde q}}^a + i {\widehat {\tilde p}}_a$, 
${\widehat a}^{\dagger}_a = {\widehat {\tilde q}}^a - 
i {\widehat {\tilde p}}_a$ for 
each pair of canonical operators 
${\widehat {\tilde q}}^a$, and ${\widehat {\tilde p}}_a$ because the number 
of these operators is even. ${\widehat a}_a$, 
${\widehat a}^{\dagger}_a$ can help in the construction of ${\cal H}_r$.

With the before machinery, the quantum version of the evolving constants is
\begin{eqnarray}
{\widehat q}^i & = & q^i (t^m (q^i , p_i ); {\widehat {\tilde q}}^a, 
{\widehat {\tilde p}}_a ) \, , \label{QECI}\\
{\widehat p}_i & = & p_i (t^m (q^i , p_i ); {\widehat {\tilde q}}^a , 
{\widehat {\tilde p}}_a ) \, \label{QECII},
\end{eqnarray}
or, equivalently,
\begin{eqnarray}
\langle \psi \mid {\widehat q}^i \mid \psi \rangle & = & 
\langle \psi \mid q^i (t^m (q^i , p_i ); {\widehat {\tilde q}}^a, 
{\widehat {\tilde p}}_a ) \mid \psi \rangle \, , \\
\langle \psi \mid {\widehat p}_i \mid \psi \rangle & = & \langle \psi \mid 
p_i ( t^m (q^i , p_i ); {\widehat {\tilde q}}^a, 
{\widehat {\tilde p}}_a ) \mid \psi \rangle \, ,
\end{eqnarray}
where the mean values are computed with the inner product 
(\ref{inner product}). In the case of parameterized systems, last equations 
reduce to the standard ones which describe the evolution of the position 
and momenta operators as well as the evolution of the mean values of 
the position and momenta operators in the Heisenberg picture. Of course, the 
well-known ordering problems for the operators might appear here too.

\section{parameterized harmonic oscillator}
\subsection{Classical dynamics}
In order to make these ideas concrete, let us consider a familiar 
example: the parameterized harmonic oscillator, which action is 
\begin{eqnarray}
S= \int d \tau \left [ \frac{d x}{d \tau} p + \frac{d t}{d \tau} p_t 
- \lambda \left ( p_t + \frac{p^2}{2m} + \frac12 m \omega^2 x^2 \right )
\right ] \, .
\end{eqnarray}
The unconstrained classical space $\Gamma$ is coordinatized by the canonical 
pairs $(x,p)$, and $(t, p_t)$. By doing the variation of the action with 
respect to $x$, $p$, $t$, and $p_t$ we find the equations of motion
\begin{eqnarray}
\frac{dp}{d \tau} = -\lambda m \omega^2 x\, , \,
\frac{dx}{d \tau} = \lambda \frac{p}{m} \, , \, 
\frac{d p_t}{d \tau} = 0 \, , \, 
\frac{dt}{d \tau} = \lambda \, . \label{equations}
\end{eqnarray}
The variation of the action with respect to the Lagrange multiplier 
$\lambda$ gives the first class constraint
\begin{eqnarray}
C= p_t + \frac{p^2}{2m} + \frac12 m \omega^2 x^2 \, .
\end{eqnarray}
The classical dynamics is the unfolding of the gauge symmetry of the system 
generated by the first class constraint $C$. The gauge orbit on the 
constrained surface $C=0$ is the integral curve of the Hamiltonian vector 
field
\begin{eqnarray}
X_{dC}& = & -\frac{\partial}{\partial t} - 
\frac{p}{m} \frac{\partial}{\partial x} + 
m\omega^2 x \frac{\partial}{\partial p} \, . \label{nullvector}
\end{eqnarray}

If we have a solution $x(\tau)$, $p(\tau)$, $t(\tau)$, and $p_t (\tau)$ of 
the equations of motion (\ref{equations}), any other solution 
$x'(\tau)$, $p'(\tau)$, $t'(\tau)$, and ${p_t}' (\tau)$ can be found 
through the relations
\begin{eqnarray}
x' (\tau) & = & \cos{(\theta(\tau))} x(\tau) + 
\frac{1}{m \omega} \sin{(\theta (\tau))} p(\tau) \, ,\nonumber\\
p' (\tau) & = & -m \omega \sin{(\theta(\tau))} x(\tau) + 
\cos{(\theta (\tau))} p(\tau) \, ,\nonumber\\
t' (\tau) & = & \frac{\theta (\tau)}{\omega} + 
t(\tau) \, , \nonumber\\
{p_t}' & = & p_t \, ,
\end{eqnarray} 
that connect all the solutions, while the Lagrange multiplier transforms as
\begin{eqnarray}
\lambda' (\tau) & = & \lambda(\tau) + \frac{1}{\omega} 
{\dot \theta} (\tau) \, ,
\end{eqnarray}
in order to leave the action invariant , here 
${\dot \theta} (\tau)= \frac{d \theta (\tau)}{d \tau}$.

Let us construct the general solution in a given gauge. We choose the 
gauge $\lambda =1$. We still have one gauge fixing to impose at $\tau =0$. We 
choose $t(0) = 0$. Using the constraint equation and the solution of 
(\ref{equations}), we obtain  
\begin{eqnarray}
x (\tau) & = &  A \cos{(\omega \tau)} + B\sin{(\omega \tau)}\, ,\nonumber\\
t (\tau) & = & \tau \, , \nonumber\\
p (\tau)& = &  -m\omega A \sin{(\omega \tau)} + 
m \omega B \cos{(\omega \tau)} \, ,\nonumber\\
p_t (\tau) & = & -\frac{1}{2} m \omega^2 ( A^2 + B^2 )\, , 
\label{particular gauge}
\end{eqnarray}
where $(A, B)$ are the physical observables that coordinatize the 
{\it physical} phase space of the system, which is $R^2$. It is clear 
that $x$, $t$, $p$ are non-observables (they depend on $\tau$). 

The two physical observables $(A, B)$ can be expressed in terms of the 
phase space variables as
\begin{eqnarray}
A & = & \cos{(\omega t)}x -\frac{1}{m\omega} 
\sin{(\omega t)} p \, ,\nonumber\\
B & = & \frac{1}{m\omega} \cos{(\omega t)}p + \sin{(\omega t)}x 
\label{physicalo}
\end{eqnarray}
Notice that $A= x(t=0) \equiv x_0$, and 
$B=\frac{p(t=0)}{m\omega} \equiv \frac{p_0}{m\omega}$, i.e., the 
position $x_0$ of the harmonic oscillator when the {\it internal clock} 
measures $t=0$, and the momentum $p_0$ when the {\it internal clock} 
measures $t=0$ are (physical) {\it observables}. Moreover, Eq. 
(\ref{physicalo}) means that the precise combination of the 
position $x=X$ and the momentum $p=P$ of the harmonic oscillator, when the 
internal clock indicates $t=T$ in the form expressed by the formula 
(\ref{physicalo}) is an observable of the (composed) system: harmonic 
oscillator + internal clock. These observables have vanishing Poisson 
brackets with the first class constraint $C$ as required by the formalism of 
constrained systems. Actually, the Dirac method requires observables to have 
weakly vanishing Poisson brackets with the first class constraints. Here, the 
observables $A$, $B$ have strong vanishing Poisson brackets with the 
constraint $C$. The Poisson brackets between $A$ and $B$ in the physical 
phase space reads
\begin{eqnarray}
\{ A , B \} & = & \frac{1}{m \omega}\, .
\end{eqnarray}

{\it Classical evolving constants}. From (\ref{particular gauge}), we 
obtain the evolving constant 
\begin{eqnarray}
x & = &  x_0 \cos{(\omega t)} + 
\frac{p_0}{m\omega} \sin{(\omega t)}\, , \label{evolving}
\end{eqnarray}  
of the system. As before mentioned, last equation admits two, related, 
interpretations. First, for any fixed point $(A,B)$ (equivalently 
$(x_0, p_0)$) of the physical phase space, (\ref{evolving}) gives the 
{\it relative evolution} of the configuration variables $x$, and $t$ of the 
system
\begin{eqnarray}
x & = & X(t;x_0,p_0)= x_0 \cos{(\omega t)} + 
\frac{p_0}{m\omega} \sin{(\omega t)} \, . \label{evolvingII}
\end{eqnarray}
Second, for any fixed $t$, it gives a one-parameter family of physical 
observables, $t$ being the parameter, on the physical phase space.  

{\it Generation of evolving constants.} We define the function $E^1$ on 
$\Gamma$ 
\begin{eqnarray}
E^1(x,t,p, p_t) & := & x - x_0 \cos{(\omega t)} - 
\frac{p_0}{m\omega} \sin{(\omega t)} \, .
\end{eqnarray}
The restriction of this function on the constraint surface is 
$E^1 \mid_{C}=0$. The action of the Hamiltonian vector field $X_{dC}$ on 
$E^1$ is   
\begin{eqnarray}
X_{dC}(E^1) & =: & E^2 = -\omega x_0 \sin{(\omega t)} + \frac{p_0}{m} 
\cos{(\omega t)} -\frac{p}{m}\, ,
\end{eqnarray}
and the restriction of $E^2$ on the constraint surface is
${E^2 \mid_{C}}=0$, and more important, the equation $E^2 \mid_{C}=0$ 
is precisely an evolving constant
\begin{eqnarray}
p & =& -m \omega x_0 \sin{(\omega t)} + 
p_0 \cos{(\omega t)}\, . \label{evolvingIII}
\end{eqnarray}
Note also that 
the action of $X_{dC}$ on $E^2$ gives again $E^1$, and the process 
ends. In other words, the evolving constant (\ref{evolvingIII}) was obtained 
from the application of the Hamiltonian vector field 
$X_{dC}$ on $E^1$, and viceversa. 

{\it The full solution}. In the present case the constraint surface is 
coordinatized by the coordinates of the physical phase space 
$(x_0 ,p_0 )$ and by the internal time $t$. Therefore, Eqs. (\ref{ECI}), and 
(\ref{ECII}) acquire the form
\begin{eqnarray}
x & = & X(t; x_0 , p_0) = x_0 \cos{(\omega t)} + \frac{p_0}{m\omega} 
\sin{(\omega t)}\, ,\nonumber\\
t & = & T(t; x_0 ,p_0 ) = t \nonumber\\
p & = & P (t; x_0 , p_0 ) = -m \omega x_0 \sin{(\omega t)} + 
p_0 \cos{(\omega t )} \, ,\nonumber\\
p_t & = & P_T (t; x_0, p_0)  = -\frac{{p_0}^2}{2m}- 
\frac{1}{2} m \omega^2 {x_0}^2 \, ,
\end{eqnarray}
Of course, last equations are also (\ref{FSI}), and (\ref{FSII}). Notice 
that Eqs. (\ref{OI}), and (\ref{OII}) acquire the form    
\begin{eqnarray}
x_0 & = & \cos{(\omega t)}x -\frac{1}{m\omega} 
\sin{(\omega t)} p \, ,\nonumber\\
p_0 & = & \cos{(\omega t)}p + m \omega  \sin{(\omega t)}x \, ,
\end{eqnarray}
and the dependence of the orbit coordinate $x^1= t$, see (\ref{GI}), is 
\begin{eqnarray}
t=T(x,t, p ,p_t) & = & t\, .
\end{eqnarray}
Last equations constitute the full solution of the 
classical dynamics of the system. 

Notice that the internal time variable $x^1=t=T(x,t,p,p_t)=t $ is not a 
physical observable because the Poisson bracket with the first class 
constraint does not vanish. Nevertheless, when we take the full 
solution into account we can express $t={\widetilde T}(x,p,p_0,x_0)$, given 
by
\begin{eqnarray}
\cos{\omega t} & =  &
\frac{\left ( \frac12 m\omega x x_0 + \frac{1}{2m} p p_0 \right )}{H_0} \, ,
\end{eqnarray}
with $H_0 = \frac{1}{2m} {p_0}^2+ \frac12 m\omega {x_0}^2 $. The above 
expression is an evolving constant. From this point of view, the 
internal clock $t$ defines a two-parameter family of physical observables 
on the physical phase space; $x$, and $p$ being the parameters. So, the 
internal clock $t$ becomes a {\it physical clock} $t(x,p)$, namely, a 
physical observable when the full solution is considered. We restrict the 
analysis to a branch of the above multivalued function to compute the 
time $t(x,p)$ at which the particle reaches the position $x$ and the 
momentum $p$ evolving from an initial position $x_0$ and momentum $p_0$
\begin{eqnarray}
t(x,p)  & = &  {\widetilde T}(x,p;x_0,p_0) = 
\frac{1}{\omega} \mbox{arc cos} \left (
\frac{\frac12 m\omega x x_0 + \frac{1}{2m} p p_0}{H_0}
 \right) \, . \label{time operator}
\end{eqnarray}
Or in terms of $x$ only 
\begin{eqnarray}
& & t_{\pm}(x)  =  \frac{1}{\omega} \mbox{arc cos} \left (
\frac{\frac12 m\omega x x_0 \pm \sqrt{\frac{1}{2m} \left ( H_0 -
\frac{1}{2}m \omega x^2\right )} p_0}{H_0} \right) \, . 
\nonumber \\
\end{eqnarray}
These classical expressions have a quantum version as we will see later. 

\subsection{Quantum dynamics}
At quantum level, as Dirac showed, the {\it physical states} are those 
killed by the first class constraint. We associate {\it abstract} operators 
with the classical coordinates and momenta, given by
\begin{eqnarray}
x \rightarrow  {\widehat X} \quad , \quad 
t \rightarrow {\widehat T} \quad, \quad
p \rightarrow {\widehat P} \quad , \quad
p_t \rightarrow {\widehat P_T} \, ,
\end{eqnarray}
which satisfy the Dirac rule
\begin{eqnarray}
{[{\widehat X}\, , \, {\widehat P} ]} =  i \hbar \quad , \quad
{[{\widehat T}\, , \, {\widehat P_T} ]} = i\hbar \, ,
\end{eqnarray}
and by inserting these operators in the quantum constraint 
${\widehat C} \mid \psi \rangle$, this equation becomes  
\begin{eqnarray}
\left ( {\widehat P_T} + 
\frac{{{\widehat P}}^2}{2m}+\frac12 m \omega^2 {\widehat X}^2 \right )
\mid \psi \rangle & = & 0 \, .
\end{eqnarray}
Any {\it physical state} can be expressed in terms of the single 
quantum number of the harmonic oscillator, in abstract Dirac notation 
\begin{eqnarray}
\mid \psi \rangle & = & \sum_n C_n \mid n \rangle \, , \nonumber\\
{\widehat I} & = & \sum_n \mid n \rangle \langle n \mid \, . 
\label{harmonic}
\end{eqnarray}
In last expression, the physical states $\mid \psi \rangle $ are 
`frozen' (i.e. they are abstract vectors), the complex coefficients $C_n$ 
are constants. Notice that we have not choosen the coordinate basis 
yet. Taking a `coordinate representation' $\mid x, t\rangle$ where the 
operators acquire the form 
\begin{eqnarray}
\langle x , t \mid {\widehat X} \mid \psi \rangle & = & x 
\langle x , t \mid \psi \rangle \, , \nonumber\\
\langle x , t \mid {\widehat T} \mid \psi \rangle & = & t \langle x , t 
\mid \psi \rangle\, , \nonumber\\
\langle x , t \mid {\widehat P} \mid \psi \rangle & = & \frac{\hbar}{i} 
\frac{\partial}{\partial x} \langle x , t \mid \psi \rangle \, ,\nonumber\\
\langle x , t \mid {\widehat P_T} \mid \psi \rangle & = & \frac{\hbar}{i} 
\frac{\partial}{\partial t} \langle x , t \mid \psi \rangle \, ,
\end{eqnarray}
any physical state vector $\mid \psi \rangle$ is expanded in the coordinate 
basis  $\mid x , t \rangle$ as
\begin{eqnarray}
\langle x,t \mid \psi \rangle & = & \psi (x,t) =
\sum_n C_n \langle x,t \mid n \rangle \, , \nonumber\\
\langle x' , t' \mid x , t \rangle & = & \sum _n 
\langle x' , t' \mid n \rangle \langle n \mid x , t \rangle \, ,
\end{eqnarray}
with $\langle x , t \mid n \rangle = e^{-\frac{i}{\hbar} E_n t} f_n (x)$, 
$E_n = \hbar \omega \left ( n + \frac12 \right )$. Thus, in the 
Dirac framework, the coordinate representation is nothing but the 
`Heisenberg picture' of the standard quantum mechanics, where the 
coordinate basis $\mid x , t \rangle$ is `rotating' and the physical 
state $\mid \psi \rangle$ is {\it fixed} (see Eq. (\ref{harmonic}) where 
the coefficients $C_n$ are {\it constant} complex numbers). 

{\it Schr\"odinger equation}. In addition, we 
can build a `Schr\"odinger basis' from the `Heisenberg basis' 
$\mid x , t \rangle$. In this `Schr\"odinger basis', which we denote 
by $\mid x \rangle$, the state vector is `moving around' the `fixed basis' 
$\mid x \rangle$. Explicitly, 
\begin{eqnarray}
\psi (x,t) & = & \langle x \mid \psi (t) \rangle \quad , \quad 
\mbox{Schr\"odinger basis $\mid x \rangle$}\, ,
\end{eqnarray}
with
\begin{eqnarray}
\mid \psi (t) \rangle & = & \sum_n {\widetilde C}_n(t) 
\mid {\widetilde n} \rangle = \sum_n C_n e^{-\frac{i}{\hbar} E_n t} 
\mid {\widetilde n} \rangle \, , \nonumber\\
\langle x \mid {\widetilde n} \rangle & = & f_n (x) \, . 
\end{eqnarray}
Taking the derivative with respect to the coordinate $t$ of $\mid \psi (t) 
\rangle$, the familiar Schr\"odinger equation emerges in the formalism 
\cite{Kuchar}
\begin{eqnarray}
i \hbar \frac{d}{d t} \mid \psi(t) \rangle & = & {\widehat H} 
\mid \psi(t) \rangle \, ,
\end{eqnarray}
with ${\widehat H} = \frac{1}{2m}{\widehat P}^2 +\frac12 m \omega^2 
{\widehat X}^2$. As usual, the physical vector $\mid \psi (t) \rangle$ 
evolves in $t$ while the coordinate basis $\mid x \rangle$ is fixed. In 
other words, if we consider the system composed of the harmonic oscillator 
{\it plus} the clock together, we are describing the evolution of the 
degrees of freedom of the harmonic oscillator with respect to the internal 
clock itself, that is to say, the evolution of one part of the system with 
respect to the rest of it. In the next section, we will carry out the same 
procedure we applied here in order to analyze the meaning of evolution in 
generally covariant quantum theories.   

{\it Quantum evolving constants.} Let us now go to the quantum version of the 
evolving constants. The Hilbert space is built with the implementation of the 
physical state vectors  
$\mid \psi \rangle = \sum_n C_n \mid n \rangle$ in the reduced Hilbert 
space ${\cal H}_{r}$ associated with the physical phase space of the 
harmonic oscillator. In the present case
\begin{eqnarray}
\psi (x_0) & = & \langle x_0 \mid \psi \rangle = \sum_n C_n f_n (x_0)\, . 
\end{eqnarray}
The inner product in ${\cal H}_{r}$
\begin{eqnarray}
\langle \psi \mid \phi  \rangle &= & 
\int d\mu(x_0) \,\, \psi^{\ast} (x_0) \phi (x_0)
\end{eqnarray}
can be determined with the condition that the operators ${\widehat x}_0$, and 
${\widehat p}_0$ be hermitian operators. 

Thus, in the classical expression 
\begin{eqnarray}
x & = & X(t;x_0,p_0)=  x_0 \cos{(\omega t)} + 
\frac{p_0}{m\omega} \sin{(\omega t)}\, ,
\end{eqnarray}
$x_0$, and $p_0$ are physical observables given by 
the Eq. (\ref{physicalo}) and they become operators acting on 
${\cal H}_r$, so the quantum version of the classical evolving constant is 
\begin{eqnarray}
{\widehat x} (t) & = & x(t; {\widehat x}_0, {\widehat p}_0) = 
{\widehat x}_0 \cos{(\omega t)} + 
\frac{{\widehat p}_0}{m\omega} \sin{(\omega t)}\, , \label{qevolvingII}
\end{eqnarray}
which is the well-known evolution equation for the position operator 
${\widehat X}$ in the Heisenberg picture. 

In addition, the classical expression 
\begin{eqnarray}
p & = & P(t;x_0,p_0) = -m\omega x_0 \sin{(\omega t)} + 
p_0 \cos{(\omega t)} \, , 
\end{eqnarray} 
has its quantum analog
\begin{eqnarray}
{\widehat p} (t) & = & p(t;{\widehat x}_0 , {\widehat p}_0)= 
-m\omega {\widehat x}_0 \sin{(\omega t)} + 
{\widehat p}_0 \cos{(\omega t)} \, , \label{qevolvingIII}
\end{eqnarray} 
and finally
\begin{eqnarray}
{\widehat p}_t & = & p_t (t; {\widehat x}_0, {\widehat p}_0)  = 
-\frac{{{\widehat p}_0}^2}{2m}- \frac{1}{2} m \omega^2 {{\widehat x}_0}^2 \, .
\end{eqnarray}
In summary, for the case of parameterized systems, the quantum 
version of the evolving constants equations constitutes the Heisenberg 
equations for the physical operators involved in each particular 
theory. In the case of the harmonic oscillator, Eqs. (\ref{qevolvingII}) 
and (\ref{qevolvingIII}).

{\it Time operator}. The classical expression (\ref{time operator}) becomes 
an operator ${\widehat T} (X,P) = t(X,P; {\widehat x}_0, {\widehat p}_0)$ 
which is defined on the reduced Hilbert space ${\cal H}_r$. Taken 
arbitrarily the order of the operators, we have
\begin{eqnarray}
{\widehat T}(X,P) & = & \frac{1}{\omega} \mbox{arc cos} \left (
\frac{\frac12 m\omega X {\widehat x}_0 + \frac{1}{2m} P {\widehat p}_0}
{{\widehat H}_0}
 \right) \, ,
\end{eqnarray}
with ${\widehat H}_0 = \frac{1}{2m} {{\widehat p}_0}^2 +
\frac12 m\omega {{\widehat x}_0}^2$. From this operator, we can compute 
the `time of arrival' operator ${\widehat T}(X)$
\begin{eqnarray}
& & {\widehat T}_{\pm}(X) = \frac{1}{\omega} \mbox{arc cos} \left (
\frac{\frac12 m\omega X {\widehat x}_0 \pm 
\sqrt{\frac{1}{2m} \left ( {\widehat H}_0 - 
\frac{1}{2}m \omega X^2\right )} {\widehat p}_0}{{\widehat H}_0} 
\right) \, , \nonumber \\
\end{eqnarray}
associated with the time at which the harmonic oscillator is detected with 
an apparatus located in $x=X$. The `time of arrival' operator for a free 
particle has been studied in \cite{Grot96}. The analysis of the `time of 
arrival' operator for the harmonic oscillator deserves to be studied.

\section{$SL(2,R)$ model with two Hamiltonian constraints}
\subsection{Classical dynamics}
Let us see how the relative evolution looks in a 
non familiar generally covariant model. A nonlinear generally covariant 
system with two Hamiltonian constraints and with one physical degree of 
freedom was introduced in \cite{Mon99}. This model mimics the constraint 
structure of general relativity. Here, we continue the study of this 
model. In particular, we display the full set of evolving constants required 
in its classical and quantum dynamics. Moreover, for a Schr\"odinger-like 
equation of motion arises in its quantum dynamics, we compare the meaning of 
time (evolution) in both, evolving constants and Schr\"odinger-like 
equation, viewpoints. 

First, a brief summary of its classical dynamics, for more details and 
its physical interpretation see Ref.\cite{Mon99}. The model is 
defined by the action 
\begin{eqnarray}
& S[{\vec u},{\vec v}, N, M , \lambda ] =  \frac12 
{\displaystyle \int}\ dt \left[\, N\, ({\cal D} \vec u^2 + \vec v^2)
+ M\, ({\cal D} \vec v^2 + \vec u^2)\, 
\right], &\label{Action} 
\end{eqnarray}
where
\begin{equation}
{\cal D} {\vec u}  =  \frac{1}{N} (\dot {\vec u} - 
\lambda {\vec u}), \ \ \ \ \ 
{\cal D} {\vec v}  =  \frac{1}{M} (\dot {\vec v} + 
\lambda {\vec v}); 
\end{equation}
the two Lagrangian dynamical variables ${\vec u}=(u^1,u^2)$ and 
${\vec v}=(v^1,v^2)$ are two-dimensional real vectors; $N$, $M$ 
and $\lambda$ are Lagrange multipliers.  The squares are taken in 
$R^{2}$: $\vec u^2 = \vec u\cdot \vec u = 
(u^{1})^{2}+(u^{2})^{2}$. The action can be put in the Hamiltonian form
\begin{eqnarray}
S [{\vec u}, {\vec v}, {\vec p}, {\vec \pi}, \lambda^m ] = 
\int d \tau \left [ {\dot {\vec u}} \cdot {\vec p} 
+ {\dot {\vec v}} \cdot {\vec \pi} - \lambda^m C_m \right] \, .
\end{eqnarray}
The canonical pairs that coordinatize the unconstrained classical phase space 
are $(u^1, p^1)$, $(u^2 , p^2)$, $(v^1 , {\pi}^1)$, and $(v^2 , \pi^2)$. Also 
$\lambda^1 =N$, $\lambda^2 =M$, and $\lambda^3=\lambda$. The first class 
constraints have the form
\begin{eqnarray}
C_1 & = & \frac12 \left ( {\vec p}^2 -{\vec v}^2 \right ) \, ,\nonumber\\ 
C_2 & = & \frac12 \left ( {\vec \pi}^2 -{\vec u}^2\right ) \, , \nonumber\\
C_3 & = & {\vec u} \cdot {\vec p} - {\vec v} \cdot {\vec \pi}\, ,  
\end{eqnarray} 
which algebra is isomorphic to the $sl(2,R)$ Lie algebra
\begin{eqnarray}
\{ C_1 , C_2 \} & = & C_3 \, \nonumber\\
\{ C_1 ,  C_3 \} & = & - 2 C_1 \, \nonumber\\
\{ C_2 , C_3 \} & = & 2 C_2 \, .
\end{eqnarray}
The classical dynamics is the unfolding of the gauge symmetry generated by 
the Hamiltonian vector fields
\begin{eqnarray}
X_{dC_1} & = & -{\vec p} \cdot {\vec \nabla}_u - 
{\vec v} \cdot {\vec \nabla}_{\pi}\, , \nonumber\\
X_{dC_2} & = & -{\vec \pi} \cdot {\vec \nabla}_v - 
{\vec u} \cdot {\vec \nabla}_{p}\, , \nonumber\\
X_{d C_3} & = & -{\vec u} \cdot {\vec \nabla}_u + 
{\vec v} \cdot {\vec \nabla}_{v} + {\vec p} \cdot {\vec \nabla}_p 
- {\vec \pi} \cdot {\vec \nabla}_{\pi}\, ,
\end{eqnarray}
associated with the first class constraints of the model.
 
The physical phase space can be coordinated by the points 
$(J, \phi, \epsilon ,\epsilon')$, and these physical observables have the 
following form
\begin{eqnarray}
\epsilon & = & {u^{1}p^{2}-p^{1}u^{2}\over|u^{1}p^{2}-p^{1}u^{2}|}
\, , \nonumber \\
\epsilon' & = &{\pi^{1}v^{2}-v^{1}\pi^{2}\over|\pi^{1}v^{2}-
v^{1}\pi^{2}|} \, , \nonumber \\
J & = & |u^{1}p^{2}-p^{1}u^{2}| \, , \nonumber  \\
\phi & = & \arctan{u^{1}v^{2}-p^{1}\pi^{2}\over
u^{1}v^{1}-p^{1}\pi^{1}}\, . \label{obs}
\end{eqnarray}
The Poisson brackets between $J$ and $\phi$ in the reduced phase space 
reads
\begin{eqnarray}
\{J , \phi \} = \epsilon \epsilon' \, .
\end{eqnarray}

{\it Classical evolving constants}. Finally, the relation between the 
Lagrangian variables 
$({\vec u}, {\vec v})$ and the physical states 
$(J, \phi, \epsilon ,\epsilon')$
\begin{eqnarray}
\left[ u^{1}v^{1}+\epsilon\epsilon' 
u^{2}v^{2}\right]\cos \phi
+ \left[ u^{1}v^{2}-\epsilon\epsilon' 
u^{2}v^{1}\right]\sin \phi &=& J \, , \label{evolvingSL}
\end{eqnarray}
which leads to the notion of evolving constants of the system 
\cite{Rovelli1990,Rovelli1991}. The evolving constants give 
the evolution of the Lagrangian variables of the system in a gauge 
invariant way, i.e., for any fixed physical state of the system 
$(J,\phi,\epsilon,\epsilon')$, Eq. (\ref{evolvingSL}) gives the change of one 
of the four coordinates as a function of the other three coordinates, say 
\begin{eqnarray} 
U^{1}(x,y,z; J,\phi,\epsilon, \epsilon') = {-\epsilon' x (z\cos 
\phi- y\sin \phi) + \epsilon J\over \epsilon(y\cos \phi+ z\sin \phi)} \, .
\label{com}
\end{eqnarray}
This relative evolution among the coordinates is gauge invariant. In 
addition, for any fixed $x,y,z$ last equation gives a three-parameter 
family of physical observables, the three parameters are the three 
coordinates $x,y,z$, on the physical phase space.

{\it Generation of evolving constants}. We start with the evolving constant 
(\ref{evolvingSL}), and define the evolving function $E^1$
\begin{eqnarray}
E^1 (u,v,p,\pi) & : = &  \left[ u^{1}v^{1}+\epsilon\epsilon' 
u^{2}v^{2}\right]\cos \phi + \left[ u^{1}v^{2}-\epsilon\epsilon' 
u^{2}v^{1}\right]\sin \phi - J\, .
\end{eqnarray}
The restriction of $E^1$ on the constraint surface vanishes, 
$E^1 \mid_C =0$. The action of the Hamiltonian vector field $X_{d H_1}$ on 
$E^1$ is
\begin{eqnarray}
X_{ d H_1} (E^1) =:  E^2 & = & -\left[ p_1 v^1 +\epsilon\epsilon' p_2 v^2 
\right] \cos \phi - \left[ p_1 v^2 - \epsilon\epsilon' p_2 v^1 \right]
\sin \phi \, ,
\end{eqnarray}
and the restriction of $E^2$ on the constraint surface vanishes, so 
$E^2 \mid_C =0 $ gives the evolving constant 
\begin{eqnarray}
\left[ p_1 v^1 +\epsilon\epsilon' p_2 v^2 \right] \cos \phi
+ \left[ p_1 v^2 - \epsilon\epsilon' p_2 v^1 \right]\sin \phi & = & 0 \, .
\end{eqnarray}
The action of $X_{d H_1}$ on $E^2$ gives zero, so the process ends. Now, we 
compute the action of the Hamiltonian vector field $X_{d H_2}$ on 
$E^1$
\begin{eqnarray}
X_{ d H_2} (E^1) =:  E^3 & = & 
\left[ u^1 \pi_1 +\epsilon\epsilon' u^2 \pi_2 \right] \cos \phi 
+ \left[ u^1 \pi_2 - \epsilon\epsilon' u^2 \pi_1 \right]\sin \phi 
\, , 
\end{eqnarray}
and the restriction of $E^3$ on the constraint surfaces vanishes, so 
$E^3 \mid_C =0 $ gives the evolving constant
\begin{eqnarray}
\left[ u^1 \pi_1 +\epsilon\epsilon' u^2 \pi_2 \right] \cos \phi 
+ \left[ u^1 \pi_2 - \epsilon\epsilon' u^2 \pi_1 \right]\sin \phi = 0 \, . 
\end{eqnarray}
The action of $X_{d H_2}$ on $E^3$ gives zero, so the process 
ends. Finally, the computation of the action of the Hamiltonian vector field 
$X_ {d D}$ on $E^1$
\begin{eqnarray}
X_{ d D} (E^1) =:  E^4 & = & - E^1 -J \, ,
\end{eqnarray}
so we recover the original evolving constant we start with, and no more 
evolving can be obtained from (\ref{evolvingSL}).

{\it The full solution}. Eqs. (\ref{FSI}), and (\ref{FSII}) acquire the 
form
\begin{eqnarray}
u^1 & = & U^1(u^2 , v^1 , v^2 ;J ,\phi , \epsilon ,\epsilon') = 
{-\epsilon' u^2 (v^2\cos \phi- v^2 \sin \phi) + 
\epsilon J\over \epsilon(v^1\cos \phi+ v^2\sin \phi)} \, ,\nonumber\\
u^2 & = & U^2(u^2 , v^1 , v^2 ;J ,\phi , \epsilon ,\epsilon') = 
u^2 \, ,\nonumber\\
v^1 & = & V^1(u^2 , v^1 , v^2 ;J ,\phi , \epsilon ,\epsilon') = 
v^1\, ,\nonumber\\
v^2 & = & V^2( u^2 , v^1 , v^2;J ,\phi , \epsilon ,\epsilon') = 
v^2\, , \nonumber \\
p_1 & = & P_1( u^2 , v^1 , v^2 ;J ,\phi , \epsilon ,\epsilon') = 
\epsilon' \left ( v^1 \sin{\phi} - v^2 \cos{\phi} \right )\, , \nonumber \\
p_2 & = & P_2(u^2 , v^1 , v^2 ;J ,\phi , \epsilon ,\epsilon') = 
\epsilon\left (v^1 \cos{\phi} + v^2 \sin{\phi} \right ) \, , \nonumber \\
\pi_1 & = & \Pi_1(u^2 , v^1 , v^2 ;J ,\phi , \epsilon ,\epsilon') = 
\frac{\epsilon u^2 v^1 + \epsilon' J \sin{\phi}}
{(v^1 \cos{\phi} +v^2 \sin{\phi}) } \, , \nonumber \\
\pi_2 & = & \Pi_2(u^2 , v^1 , v^2 ;J ,\phi , \epsilon ,\epsilon') = 
\frac{\epsilon u^2 v^2 - \epsilon' J \cos{\phi}}
{(v^1 \cos{\phi} +v^2 \sin{\phi}) } \, , \label{fullsolution} 
\end{eqnarray}
and the Eqs. (\ref{OI}), and (\ref{OII}) are precisely the Eqs. (\ref{obs}) 
while the Eqs. (\ref{GI}) acquire the form
\begin{eqnarray}
u^2 & = & U^2(u^i , v^i , p_i , \pi_i) = u^2 \, ,\nonumber\\
v^1 & = & V^1(u^i , v^i , p_i , \pi_i) = v^1 \, ,\nonumber\\
v^2 & = & V^2(u^i , v^i , p_i , \pi_i) = v^2 \, .
\end{eqnarray}
So, the dynamics of this model can be described in a relational fashion 
way. The difference with respect to parameterized systems, as the example of 
the harmonic oscillator previously analyzed, is that in the present case a 
single internal time variable is not enough, rather, we need three internal 
time variables. In the way we have expressed the full solution 
(\ref{fullsolution}), $u^2, v^1 , v^2$ are clocks, i.e., once the component 
$u^2$ of the position of the first particle, and the position 
$(v^1 , v^2)$ of the second particle are known, the change of the component 
$u^1$ of the first particle and the change of the momenta of both particles 
${\vec p}$, ${\vec \pi}$ are also known when the system is an particular 
physical state $(J, \phi, \epsilon ,\epsilon')$. Therefore, the 
full relational evolution of the system is expressed in terms of three 
internal clocks $u^2$, $v^1$, and $v^2$. 

\subsection{Quantum dynamics}
At quantum level, the model is characterized by the following set of 
observables
\begin{eqnarray}
{\widehat J} \mid m, \epsilon, \epsilon' \rangle & = & m \hbar 
\mid m, \epsilon, \epsilon' \rangle \, , \nonumber\\
{\widehat \epsilon} \mid m, \epsilon, \epsilon' \rangle & = & \epsilon
\mid m, \epsilon, \epsilon' \rangle \, , \nonumber\\
{\widehat \epsilon'} \mid m, \epsilon, \epsilon' \rangle & = & \epsilon'
\mid m, \epsilon, \epsilon' \rangle \, ,
\end{eqnarray}
and the physical states are given by
\begin{eqnarray}
\mid \psi \rangle = \sum_{m \epsilon,\epsilon'} 
C_{m,\epsilon,\epsilon'} \mid m,\epsilon,\epsilon' \rangle \, ,\label{state}
\end{eqnarray}
in abstract Dirac notation. In the `coordinate representation' 
$\mid u , v , \alpha , \beta \rangle$, which is nothing but the Heisenberg 
picture in standard quantum mechanics because all the coordinates 
$(u,v,\alpha,\beta)$ are put at the same level, the state reads
\begin{eqnarray}
\psi (u,v,\alpha,\beta) & = &
\langle u , v , \alpha ,\beta \mid \psi \rangle  = 
\sum_{m \epsilon,\epsilon'} C_{m,\epsilon,\epsilon'} 
\langle u , v , \alpha ,\beta \mid m,\epsilon,\epsilon' \rangle \, ,
\end{eqnarray}
with $\langle u , v , \alpha ,\beta \mid m,\epsilon,\epsilon' \rangle =
e^{im(\epsilon \alpha -\epsilon' \beta)} J_m \left ( \frac{uv}{\hbar} 
\right )$. Thus the basis $\mid u , v , \alpha , \beta \rangle$ is 
`rotating' and the state $\mid \psi \rangle $ is {\it fixed}, i.e., 
the coefficients $C_{m\epsilon\epsilon'}$ are {\it constant} complex 
numbers. The `coordinate representation' appears as the most `democratic' 
basis because it does not prefer one coordinate more than the others. 

{\it Schr\"odinger equation}. In the same sense that in parameterized 
systems we were able to build a `Schr\"odinger basis' from the Heisenberg 
basis, we can do the same here, and rewrite the physical state 
(\ref{state}). In the present example, we can build two 
Schr\"odinger bases $\mid u,v,\beta\rangle$, and 
$\mid u,v,\alpha\rangle$. In the first one, the physical state vector 
(\ref{state}) is expressed as
\begin{eqnarray}
\psi (u,v, \alpha, \beta) & = &
\langle u , v , \beta \mid \psi (\alpha) \rangle \, , 
\end{eqnarray}
with
\begin{eqnarray}
\mid \psi (\alpha) \rangle & = & \sum_{m \epsilon,\epsilon'} 
{\widetilde C}_{m,\epsilon,\epsilon'} (\alpha) 
{\widetilde{\mid m,\epsilon,\epsilon' \rangle}}  =  
\sum_{m \epsilon,\epsilon'} 
C_{m,\epsilon,\epsilon'} e^{im\epsilon \alpha}
{\widetilde{\mid m,\epsilon,\epsilon' \rangle}} \, ,\nonumber\\
\langle u,v,\beta {\widetilde{\mid m,\epsilon,\epsilon' \rangle}} & = & 
e^{-im\epsilon' \beta} J_m \left ( \frac{uv}{\hbar} \right )\, .
\end{eqnarray}
Taking the derivative with respect to the coordinate $\alpha$ of 
$\mid \psi (\alpha) \rangle$, a Schr\"odinger equation emerges in the 
formalism 
\begin{eqnarray}
i\hbar \frac{d}{d \alpha} 
\mid \psi (\alpha) \rangle & = & -\frac{\epsilon}{\epsilon'}{\widehat O}_{34}
\mid \psi (\alpha) \rangle \, , \label{alpha time}
\end{eqnarray}
and the physical observable ${\widehat O}_{34}$ has the form
\begin{eqnarray}
\langle u , v, \beta \mid {\widehat O}_{34} \mid \psi \rangle & = & 
-\frac{\hbar}{i} \frac{\partial}{\partial \beta} \langle u , v , \beta \mid 
\psi \rangle \, ,
\end{eqnarray}
in the `Schr\"odinger basis' $\mid u , v, \beta\rangle$. As expected, in the 
Schr\"odinger basis $\mid u,v,\beta\rangle$ , the state 
$\mid \psi (\alpha )\rangle$  evolves while the basis  
$\mid u,v,\beta\rangle$ is fixed with respect to $\alpha$. This is not a 
matter of terminology, in fact the evolution equation (\ref{alpha time}) 
is well defined, and we are really able of describing evolution under this 
picture, namely, to describe the change of the some part of the whole state 
with respect to rest of it, in complete agreement with the spirit of 
relationism. 

In the second Schr\"odinger basis $\mid u,v,\alpha \rangle$, the 
physical state vector (\ref{state}) is expressed as
\begin{eqnarray}
\psi (u,v, \alpha, \beta) & = &
\langle u , v , \alpha \mid \psi (\beta) \rangle \, , 
\end{eqnarray}
with
\begin{eqnarray}
\mid \psi (\beta) \rangle & = & \sum_{m \epsilon,\epsilon'} 
{\widetilde{\widetilde C}}_{m,\epsilon,\epsilon'} (\beta) 
{\widetilde{\widetilde{\mid m,\epsilon,\epsilon' \rangle}}} = 
\sum_{m \epsilon,\epsilon'} 
C_{m,\epsilon,\epsilon'} e^{-im\epsilon' \beta} 
{\widetilde{\widetilde{\mid m,\epsilon,\epsilon' \rangle}}} \, ,\nonumber\\
\langle u,v,\alpha 
{\widetilde{\widetilde{\mid m,\epsilon,\epsilon' \rangle}}} & = & 
e^{im\epsilon \alpha} J_m \left ( \frac{uv}{\hbar} \right )\, .
\end{eqnarray}
Taking the derivative with respect to the coordinate $\beta$ of 
$\mid \psi (\beta) \rangle$, a Schr\"odinger equation emerges in the 
formalism 
\begin{eqnarray}
i\hbar \frac{d}{d \beta} 
\mid \psi (\beta) \rangle & = & \frac{\epsilon'}{\epsilon} {\widehat O}_{12}
\mid \psi (\beta) \rangle \, , \label{beta time}
\end{eqnarray}
and the physical observable ${\widehat O}_{12}$ has the form
\begin{eqnarray}
\langle u , v, \alpha \mid {\widehat O}_{12} \mid \psi \rangle & = & 
\frac{\hbar}{i} \frac{\partial}{\partial \alpha} \langle u , v , \alpha \mid 
\psi \rangle \, ,
\end{eqnarray}
in the `Schr\"odinger basis' $\mid u , v, \alpha \rangle$.

{\it Quantum evolving constants}. The quantum version of the evolving 
constants is as follows. More precisely, the quantum version of the 
full classical solution ({\ref{fullsolution}) is expressed as
\begin{eqnarray}
{\widehat u}^1 & = & u^1(u^2 , v^1 , v^2 ;
{\widehat J} ,{\widehat{\sin\phi}}, {\widehat{\cos\phi}} , 
{\widehat{\epsilon}} ,{\widehat\epsilon'}) =  
\frac{-{\widehat \epsilon}' {\widehat \epsilon} ( v^2 
{\widehat{\cos{\phi}}}- v^1 {\widehat{\sin{\phi}}}) + 
{\widehat J}}{v^1 {\widehat{\cos{\phi}}}- 
v^2 {\widehat{\sin{\phi}}})} \, ,\nonumber\\
{\widehat p}_1 & = & p_1 (u^2 , v^1 , v^2 ;
{\widehat J} ,{\widehat{\sin\phi}}, {\widehat{\cos\phi}} , 
{\widehat{\epsilon}} ,{\widehat\epsilon'}) = 
{\widehat \epsilon}' \left ( v^1 {\widehat{\sin{\phi}}} - 
v^2 {\widehat{\cos{\phi}}} \right )\, , \nonumber \\
{\widehat p}_2 & = & p_2 (u^2 , v^1 , v^2 ;
{\widehat J} , {\widehat{\sin\phi}}, {\widehat{\cos\phi}} , 
{\widehat{\epsilon}} ,{\widehat\epsilon'}) = 
{\widehat \epsilon} \left ( v^1 {\widehat{\cos{\phi}}} + 
v^2 {\widehat{\sin{\phi}}} \right )\, , \nonumber \\
{\widehat \pi}_1 & = & \pi_1 (u^2 , v^1 , 
v^2 ;{\widehat J} , {\widehat{\sin\phi}}, {\widehat{\cos\phi}} , 
{\widehat{\epsilon}} ,{\widehat\epsilon'}) = 
\frac{{\widehat{\epsilon}} u^2 v^1 + 
{\widehat{\epsilon}}' {\widehat J} {\widehat{\sin{\phi}}}}
{v^1 {\widehat{\cos{\phi}}}+ 
v^2 {\widehat{\sin{\phi}}}} \, ,\nonumber\\
{\widehat \pi}_2 & = & \pi_2 (u^2 , v^1 , v^2 ;{\widehat J} ,
 {\widehat{\sin\phi}}, {\widehat{\cos\phi}}, 
{\widehat{\epsilon}} ,{\widehat\epsilon'}) = 
\frac{{\widehat{\epsilon}} u^2 v^2 -
{\widehat{\epsilon}}' {\widehat J} {\widehat{\cos{\phi}}}}
{v^1 {\widehat{\cos{\phi}}}+ 
v^2 {\widehat{\sin{\phi}}}} \, . \label{qfullsolution} 
\end{eqnarray}
The meaning of the first equation in (\ref{qfullsolution}) is the 
following: we have to take the mean value of the operator ${\widehat u}^1$ 
with respect to generic states $\mid \psi \rangle$ of the reduced Hilbert 
space ${\cal H}_r$ of the model \cite{Mon99}. 

In summary, the quantum dynamics of parameterized systems can be described 
in terms of a `Schr\"odinger equation' or in terms of the `Heisenberg 
picture'. The Schr\"odinger equation arises as a consequence of the Dirac 
quantization, as we have seen for the case of the harmonic 
oscillator. On the other hand, in the $SL(2,R)$ model, we were able to build 
two (dependent) 
Schr\"odinger equations, and thus to identify two (dependent) internal time 
variables $\alpha$ and $\beta$ with respect to which the physical states of 
the $SL(2,R)$ model evolve. This does not mean that is always possible to 
single out in general an internal time variable, given by a Schr\"odinger 
equation, in generally covariant theories once the Dirac quantization has 
been performed. Therefore, in general, a Schr\"odinger equation does not 
arise in the formalism. The Schr\"odinger picture, when this picture emerges 
in the formalism as a consequence of the Dirac quantization, singles out one 
internal clock only. More important, the quantization of generally covariant 
theories based on the reduced Hilbert space (generalized Heisenberg picture) 
need $M$ internal clocks, where $M$ is the number of first class 
constraints. In the case of the $SL(2,R)$ model, the clocks are 
$u^2$, $v^1$, and $v^2$ in the generalized Heisenberg picture. In the 
Schr\"odinger picture, the internal clock is given by $\alpha$ (or $\beta$). 

{\it Classical limit}. Now, we compare the quantum evolving constants of 
the $SL(2,R)$ model 
with those of the harmonic oscillator in order to get insights on the 
classical limit of generally covariant theories, and in particular of the 
$SL(2,R)$ model. We expect that the classical limit of generally covariant 
theories should be attached to the concept of {\it coherent states} as it 
happens in standard quantum mechanics (parameterized systems). In the case 
of the harmonic oscillator, the coherent states are roughly those states 
$\mid \psi\rangle$ in the reduced Hilbert space ${\cal H}_r$ such that the 
mean values $\langle \psi \mid {\widehat x} (t) \mid \psi \rangle$, and 
 $\langle \psi \mid {\widehat p} (t) \mid \psi \rangle$ reproduce the 
classical behavior of the system. Of course this condition is not enough to 
single out the coherent states of the system. In addition, those states have 
also to minimize the uncertainty relations of position and momentum. Of 
course, these two conditions are still not enough to identify the coherent 
states due to the fact that both conditions are satisfied by both squeezed 
and coherent states. In the case of the parameterized harmonic oscillator a 
mechanism that identifies the coherent states is available following 
standard methods. It is natural to expect that a combination of the coherent 
states approach to the quantization of generally covariant 
theories \cite{Klauder1997} with the full set of evolving constants of motion 
required in their quantum dynamics displayed here could bring the 
classical limit of constrained systems.


\section{concluding remarks}
We have displayed the full solution of the relational evolution of the 
degrees of freedom of fully constrained theories with a finite number of 
degrees of freedom (see Eqs. (\ref{FSI}), and (\ref{FSII})). Our procedure 
follows from the embedding equations of the coordinates and momenta in 
the unconstrained phase space (see Eqs. (\ref{ECI}), and (\ref{ECII})) plus 
the expressions of the $M$ internal time variables (see Eq. (\ref{GI})). The 
form of the solution containts all the evolving constants of motion needed 
in the description of the classical dynamics of fully constrained 
theories, i.e., we have given the full mathematical solution to the 
Rovelli's point of view on the `problem of time' pioneered in 
Refs. \cite{Rovelli1990,Rovelli1991a,Rovelli1991}. Of course, the physical 
(and phylosophical) interpretation is due to Rovelli. Also, we have 
explored a method to generate 
those evolving constants. This method consists in the repeated application 
of the Hamiltonian vector fields associated with the first class constraints 
on some initial evolving constant. Combining the expressions of this 
evolving constants with the expressions of the physical observables the full 
relational evolution of the coordinates and momenta is obtained. Finally, we 
have also analysed on a general setting the quantum version of the relational 
evolution of the degrees of freedom of fully constrained theories. 

To find the full solution of the relational evolution of the degrees 
of freedom for gravity, matter fields coupled to gravity (see 
\cite{Rovelli1991CQG} for the first steps), topological quantum field 
theories, or for a background-independent string theory constitutes one of 
the challenges of the new millenium.

\section*{acknowledgments}

I am indebted to Carlo Rovelli for conversations about the `problem of 
time'. I also thank financial support provided by the 
{\it Sistema Nacional de Investigadores} (SNI) of the Secretar\'{\i}a de 
Educaci\'on P\'ublica (SEP) of Mexico.


\end{document}